\documentclass[journal]{IEEEtran}

\usepackage{verbatim}  
\usepackage{vector}  
\usepackage{epsf}
\usepackage{epsfig}
\usepackage{epstopdf}
\usepackage{enumerate}
\usepackage{amsmath}
\usepackage{amssymb}
\usepackage{algorithm}
\usepackage{algpseudocode}

\usepackage{cite}

\IEEEoverridecommandlockouts

\title{Offset-Based Beamforming: A New Approach to Robust Downlink Transmission}

\author{\IEEEauthorblockN{Mostafa Medra \quad Yongwei Huang  \quad Timothy N. Davidson\\}

\thanks{This work was supported in part by the Natural Sciences and Engineering Research Council of Canada under grant RGPIN-2015-06631.}
\thanks{Preliminary versions of portions of this work appear in
\textit{Conf. Rec. 49th Asilomar Conf. Signals, Systems, Computers, Proc. 9th IEEE Signal Process. Wkshp Sensor Array and Multichannel Signal Process., and Conf. Rec. 50th Asilomar Conf. Signals, Systems, Computers}}
\thanks{Mostafa Medra was with, and Timothy N. Davidson is with, the Department of Electrical and Computer Engineering, McMaster University, Hamilton, Ontario, Canada. (E-mail: davidson@mcmaster.ca). Mostafa Medra is now with The Edward S. Rogers Sr. Department of Electrical and Computer Engineering, University of Toronto, Toronto, Ontario, Canada. (E-mail: mmedra@ece.utoronto.ca). Yongwei Huang is with the School of Information Engineering, Guangdong University of Technology, University Town, Guangzhou 510006, China. (E-mail: ywhuang@gdut.edu.cn).}

}

\begin{document}

\maketitle

\begin{abstract}

The design of a set of beamformers for the multi-user multiple-input single-output (MISO) downlink that provides the receivers with prespecified levels of quality-of-service (QoS) can be quite challenging when the channel state information is not perfectly known at the base station. The constraint of having the SINR meet or exceed a given threshold with high probability is intractable in general, which results in problems that are fundamentally hard to solve. In this paper, we will develop a high-quality approximation of the SINR outage constraint that, along with a semidefinite relaxation, enables us to formulate the beamformer design problem as a convex optimization problem that can be efficiently solved. For systems in which the uncertainty size is small, a further approximation yields algorithms based on iterative evaluations of closed-form expressions that have substantially lower computational cost. Since finding the beamforming directions incurs most of the computational load of these algorithms, analogous power loading algorithms for predefined beamforming directions are developed and their performance is shown to be close to optimal.  When the system contains a large number of antennas, the proposed power loading can be obtained at a computational cost that grows only linearly in the number of antennas. The proposed power loading algorithm provides an explicit relationship between the outage probability required and the power consumed, which allows us to precisely control the power consumption, and automatically  identifies users who are consuming most of the power resources. The flexibility of the proposed approach is illustrated by developing a power loading technique that minimizes an average notion of outage.


\end{abstract}


\section{Introduction}

The directional signalling capabilities of  base stations (BSs) that have multiple transmit antennas enable a variety of techniques~\cite{SymbollevelandMulticast} for simultaneously transmitting independent messages to multiple single-antenna receivers, including dirty paper coding~\cite{TheCapacityRegion}, vector perturbation precoding~\cite{Avectorperturbationtechnique2}, lattice reduction precoding~\cite{latticereductionaided}, Tomlinson-Harashima precoding~\cite{Precodinginmultiantenna}, rate splitting~\cite{RobustTransmissioninDownlink}, per-symbol beamforming~\cite{ConstructiveMultiuserInterference}, and conventional linear beamforming~\cite{ShiftingtheMIMO}. Of these signalling techniques, conventional linear beamforming has the simplest implementation and will be the focus of this paper. In particular, we will consider scenarios  in which the users that have been scheduled  for transmission  specify the quality-of-service (QoS) that they expect to receive. In that setting, the BS designs the set of beamformers to ensure that the signal-to-interference-and-noise ratio (SINR) at each receiver meets the target level that is implicitly specified by that user's QoS requirements. When the BS has perfect knowledge of the channel to each user, the beamformers that minimize the total transmitted power required to achieve the SINR targets can be efficiently found \cite{Jointoptimal,Reference2,Solutionofthemultiuser,OptimalMultiuserTransmit}. However, in practice these channels are estimated and possibly predicted. In time division duplexing (TDD) systems the estimation is typically performed during the training phase on the uplink, whereas in frequency division duplexing (FDD) systems, each receiver estimates its channel and feeds back a quantized version of that estimate to the BS. Since the BS  has only estimates of the users' channels, it can only estimate the receivers' SINRs. Those estimates are, quite naturally, uncertain and hence there is a possibility that a design performed using the estimated channels will fail to meet the SINR targets when the beamformers are implemented.

A prominent approach to designing a precoder that can control the consequent outage is to postulate a model for the uncertainty in the channel estimates and to seek designs that control the outage probability under that uncertainty model. In some cases the approach involves jointly designing the beamforming directions and the power allocated to these directions (e.g.,~\cite{Optimalpowercontrol,Probabilisticallyconstrained,OutageConstrained,LowComplexityRobustMISO}), while in other cases the beamforming directions are designed based on the channel estimates only, and the uncertainty model is incorporated into the design of the power loading; e.g., \cite{Coordinateupdate,Arobustmaximin,ATractableMethod}. Unfortunately, in most settings the outage constraint has proven to be intractable (an exception is the case in \cite{Coordinateupdate}), and hence the goal has been to develop computationally efficient algorithms that can  manage the outage probability. One possible strategy for doing so is to seek ``safe" approximations of the robust optimization problem \cite{Reference4}. When such approximations result in a feasible design problem, the solution is guaranteed to satisfy the constraints of the original problem, but these approximations can be quite conservative; e.g., \cite{OutageConstrained,Probabilisticallyconstrained}. An alternative strategy is to develop approximations of the outage constraint that typically provide good performance, but might not necessarily guarantee that their solution is feasible for the original problem; e.g.,  \cite{LowComplexityRobustMISO,Optimalpowercontrol}. The approach taken in this paper falls into that class.

The development of the proposed offset-based approach begins with the rewriting of the SINR constraint as the non-negativity of a random variable. That random variable is a non-convex quadratic function of the uncertainties, in which the quadratic kernel is a quartic function of the beamformers. Then, we approximate the non-negativity constraint on the random variable by the constraint that its mean is  larger than a given multiple of its standard deviation. For the case of Gaussian channel uncertainties, the mean and  standard deviation are quadratic and quartic function of the beamformers, respectively. That fact enables the application of semidefinite relaxation techniques to obtain a convex formulation of (a relaxed version of) the approximated problem.  While that design technique is quite effective, the computational cost of solving the convex conic program with semidefinite constraints is significant. By making a further approximation that is suitable for systems with reasonably small uncertainties, we obtain a design formulation for which the KKT optimality conditions have a simpler structure. That simpler structure facilitates the development of an approximate solution method that only requires the iterative evaluation of closed-form expressions. Further approximations reveal a connection with the low-complexity technique developed in~\cite{LowComplexityRobustMISO}.

An analysis of the computational cost of these precoder design techniques shows that it is the calculation of the beamforming directions that consumes most of the required computational resources, and that when these directions are defined in-advance, the computational load can be significantly reduced. Accordingly, we develop variants of our precoder design algorithms that perform power loading on a set of fixed beamforming directions. These algorithms have low computational costs, and provide performance that is close to that of the optimal power loading algorithm~\cite{Coordinateupdate}. Furthermore, for systems with a large number of antennas (i.e., ``massive MIMO") in which the channel hardens, we develop a variant of our power loading algorithm that has a computational cost that grows only linearly with the number of antennas.

In practice, the BS has limited power available for transmission, and it is possible that the power required to serve the scheduled users with the required outage probabilities may exceed that limit. In some of these scenarios, some users suffer from a weak channel, or from having their channels closely aligned with those of other users. When that happens, such users consume most of the power transmitted by the BS. This suggests opportunities to reschedule users. On the other hand, some users might be close to the BS and experiencing a relatively strong channel; a case that suggests opportunities for doing some sort of power saving. The proposed power loading algorithm provides an explicit relationship between the required outage probabilities and the consumed power, which allows us to address these issues. Using this explicit power-outage relationship we can reduce the required power when the resulting increases in the outage probabilities are tolerable, and we can identify users that consume  excessive amounts of power.

The above-mentioned designs are ``fair" in the sense that they seek to provide each user with their specified outage probability. However, the proposed design techniques are quite flexible, and can accommodate other objectives, such as the sum of the outage probabilities. As we will demonstrate, such designs can improve the average performance of the users.

\section{System model}

We consider a scenario in which a BS that has $N_t$ antennas communicates with $K$ single-antenna users over a narrow-band channel. In the linear beamforming transmission case, the transmitted signal can be written as $\mathbf{x}= \sum_{k=1}^K\mathbf{w}_k s_k,$ where $s_k$ is the normalized data symbol intended for user $k$, and $\mathbf{w}_k$ is the associated beamformer vector. For later reference we let $\mathbf{u}_k =\mathbf{w}_k /\|\mathbf{w}_k \| $ denote the beamforming direction for user $k$, and let $\beta_k=\|\mathbf{w}_k \|^2$ denote the power allocated to that direction. Hence, $\mathbf{w}_k= \sqrt{\beta_k}\mathbf{u}_k $. The received signal at user $k$ is modelled as
\begin{equation}\label{rcvd_sig}
    y_k= \mathbf{h}_k^H \mathbf{w}_k s_k + \textstyle\sum_{j \neq k}\mathbf{h}_k^H \mathbf{w}_j s_j + n_k,
\end{equation}
where $\mathbf{h}_k^H$ is the vector of complex channel gains between the antennas at the BS and user $k$, and $n_k$ is the additive zero-mean circular complex Gaussian noise at that user.
Under this model, if we let $\sigma_k^2$ denote the noise variance, then the SINR at user $k$ is
\begin{equation}\
    \text{SINR}_k= \frac{| \mathbf{h}_k^H \mathbf{w}_k|^2}{\sum_{j \neq k} | \mathbf{h}_k^H  \mathbf{w}_j|^2 + \sigma_k^2}.
\end{equation}

The design of a set of beamformers  $\{ \mathbf{w}_k\}_{k=1}^K$ so that the SINRs satisfy specified target  values (i.e., $\text{SINR}_k\geq \gamma_k$) requires the knowledge of the channel vectors $\{\mathbf{h}_k\}_{k=1}^K$. However, the BS  has only estimates of $\{\mathbf{h}_k\}_{k=1}^K$, and hence its estimates of the SINRs at the receivers are uncertain. Accordingly, we will incorporate the channel uncertainty model into the design process. In particular, we will consider systems in which the uncertainty can be modelled using the simple additive model,
\begin{equation}\label{uncertainty}
   \mathbf{h}_k= \mathbf{h}_{e_k} +\mathbf{e}_k,
\end{equation}
where $\mathbf{h}_{e_k}$ is the BS's estimate of the channel to user $k$, and the uncertainty in that estimate is characterized by the distribution of the elements of $\mathbf{e}_k$.
In this paper, we will focus on scenarios in which $\mathbf{e}_k$ can be modelled as a circular complex Gaussian random variable  with mean $\mathbf{m}_k$ and covariance $\mathbf{C}_k$; i.e., $\mathbf{e}_k \backsim \mathcal{CN} (\mathbf{m}_k, \mathbf{C}_k)$. One scenario in which that model is applicable is that of a TDD scheme operating in a slow fading environment, in which the BS estimates the channel on the uplink using a linear estimator and exploits channel reciprocity. When the channel gains are uncorrelated and the BS employs the best linear unbiased estimator (BLUE), $\mathbf{e}_k \backsim \mathcal{CN} (0, \sigma_{e_k}^2\mathbf{I})$, and we will pay particular attention to that case. (Robust beamforming schemes for uncertainty models tailored to the FDD case were developed in  \cite{LowComplexityRobustMISO}.)

Now if we let $\delta_k$ denote the maximum tolerable outage probability for user $k$, the generic joint beamforming and power loading problem can be written as
\begin{subequations}\label{outage_min}
\begin{align}
    \min_{\substack{\mathbf{w}_k}}  \quad  &\textstyle\sum_{k=1}^K \mathbf{w}_k^H \mathbf{w}_k \\
   \text{subject to} \quad  & \text{Prob}(\text{SINR}_k \geq \gamma_k)\geq 1- \delta_k, \quad \forall k. \label{sinr5}
     \end{align}
\end{subequations}
This problem is hard to solve due to the intractable probabilistic outage constraint in \eqref{sinr5} even when the uncertainty is Gaussian~\cite{Optimalpowercontrol,Probabilisticallyconstrained,OutageConstrained}.
In order to resolve that intractability, a variety of approximations of the problem in \eqref{outage_min} by problems that are tractable have been proposed \cite{Optimalpowercontrol,Probabilisticallyconstrained,OutageConstrained,LowComplexityRobustMISO}.
In many cases, the class of approximations that is considered is restricted to the class of ``safe'' approximations~\cite{Reference4}.
Such approximations are structured so that they guarantee that any solution of the approximate problem is feasible for the original problem
in \eqref{outage_min}. However, in the downlink beamforming application, such approximations can be quite conservative, in the sense that the
feasible set of the approximate problem is significantly smaller than that of the original problem; cf. \eqref{outage_min}. That can result in
instances of the approximate problem being infeasible when the original problem has a solution, or in beamformer designs that consume significantly more power than necessary. The approximation that we will develop below is not structurally constrained in this way, but it typically performs well in practice. Furthermore, its simple form provides considerable flexibility in its application, and facilitates the development of highly-efficient algorithms.
%

\section{Principles of the offset-based approach}\label{sect3}

The derivation of the proposed approximation of the outage probability begins by rewriting $\text{SINR}_k \geq \gamma_k$ as $\mathbf{h}_k^H \mathbf{Q}_k \mathbf{h}_k - \sigma_k^2 \geq 0$, where
\begin{equation}
\begin{aligned}
    \mathbf{Q}_k &= \mathbf{w}_k \mathbf{w}_k^H/\gamma_k-\textstyle\sum_{j \neq k} \mathbf{w}_j \mathbf{w}_j^H \\
                 &= \beta_k  \mathbf{u}_k \mathbf{u}_k^H/\gamma_k-\textstyle\sum_{j \neq k} \beta_j \mathbf{u}_j \mathbf{u}_j^H.
     \end{aligned}
\end{equation}
That is, the probability that $\text{SINR}_k \geq \gamma_k$ is the same as the probability that the term $\mathbf{h}_k^H \mathbf{Q}_k \mathbf{h}_k - \sigma_k^2$ is non-negative. Under the additive  uncertainty model in \eqref{uncertainty}, we observe that $\mathbf{h}_k^H \mathbf{Q}_k \mathbf{h}_k - \sigma_k^2$ is an indefinite quadratic function of the uncertainty, $\mathbf{e}_k$. In particular, we can formulate the SINR constraint as follows
\begin{equation}\label{SINR_reformulation}
   f_k(\mathbf{e}_k)=\mathbf{h}_{e_k}^H \mathbf{Q}_k \mathbf{h}_{e_k} + 2 \text{Re}(\mathbf{e}_k^H  \mathbf{Q}_k \mathbf{h}_{e_k}  ) + \mathbf{e}_k^H  \mathbf{Q}_k \mathbf{e}_k  - \sigma_k^2 \geq 0.
\end{equation}

The key observation that underlies the offset approximation is that for uncertainties $\mathbf{e}_k$ that are reasonably concentrated,  if we design the beamforming vectors so that the mean value of  $f_k(\mathbf{e}_k)$, denoted by $\mu_{f_k}$, is a significant multiple of its standard deviation, denoted by $\sigma_{f_k}$, then that user will achieve a low outage probability. If we let $r_k$ denote that multiple for the $k$th user, then the resulting approximation of the SINR constraint, $\text{Prob}(\text{SINR}_k \geq \gamma_k)\geq 1- \delta_k$,  can be written as
\begin{equation}\label{offset_constr}
   \mu_{f_k} \geq r_k \sigma_{f_k}.
\end{equation}
In order to develop an intuitive rationale for that approximation for the outage probability, we observe that when $\mathbf{e}_k$ in \eqref{uncertainty} is Gaussian, $f_k(\mathbf{e}_k)$  has a generalized chi-square distribution \cite{OntheDistributionof}. We also observe that the term that complicates the calculation  of the relevant tail probability (i.e., Prob ($f_k(\mathbf{e}_k)<0$)) is the indefinite quadratic term $\mathbf{e}_k^H  \mathbf{Q}_k \mathbf{e}_k$ in \eqref{SINR_reformulation}. To have reasonable outage performance, the norm of the channel uncertainty $\mathbf{e}_k$ in \eqref{uncertainty} should be relatively small compared to the norm of the channel; cf., \cite{MIMObroadcast}. In that case, the constant and linear terms in  \eqref{SINR_reformulation} will tend to dominate the quadratic term. Furthermore, the distribution of $ \mathbf{e}_k^H  \mathbf{Q}_k \mathbf{e}_k$  is ``bell shaped" since $\mathbf{Q}_k$ generically has one positive and $K-1$ negative eigenvalues. Now if we approximate the quadratic term $\mathbf{e}_k^H  \mathbf{Q}_k \mathbf{e}_k$ by a Gaussian term of the same mean and  variance, then the distribution of $f_k(\mathbf{e}_k)$ becomes Gaussian and the constraint in \eqref{offset_constr} provides precise control over the tail probability.
In other words, the constraint in \eqref{offset_constr} provides precise control of the tail probability of the Gaussian approximation of $f_k(\mathbf{e}_k)$. These insights, and the guidance that they provide on the choice of $r_k$, are discussed in more detail in Appendix~\ref{r_value_sel}.

To be able to use the offset approximation in \eqref{offset_constr} in a low-complexity design algorithm, we need to obtain expressions for  $\mu_{f_k}$ and $ \sigma_{f_k}$ in terms of the design variables $\mathbf{w}_k= \sqrt{\beta_k}\mathbf{u}_k$. As shown in Appendix \ref{mean_var_der}, when  $\mathbf{e}_k \backsim \mathcal{CN} (\mathbf{m}_k, \mathbf{C}_k)$
\begin{subequations}\label{mean_eqn_c}
\begin{align}
  \mu_{f_k}& = \mathbb{E} \{f_k(\mathbf{e}_k)\} \nonumber \\
&= (\mathbf{h}_{e_k}+\mathbf{m}_k)^H \mathbf{Q}_k (\mathbf{h}_{e_k}+\mathbf{m}_k) - \sigma_k^2 + \mathbf{w}_k^H \mathbf{C}_k \mathbf{w}_k /\gamma_k \nonumber\\
 & \qquad -\sum_{j \neq k} \mathbf{w}_j^H \mathbf{C}_k  \mathbf{w}_j^H, \\
\sigma_{f_k}^2 & = \text{var} \{f_k(\mathbf{e}_k)\} \nonumber \\
&=  2 (\mathbf{h}_{e_k}+\mathbf{m}_k)^H   \mathbf{C}_k^{1/2} \mathbf{Q}_{k}^2 \mathbf{C}_k^{1/2} (\mathbf{h}_{e_k}+\mathbf{m}_k) \nonumber \\
 &\qquad +\text{tr} (\mathbf{C}_k^{1/2} \mathbf{Q}_{k} \mathbf{C}_k^{1/2} )^2.
    \end{align}
\end{subequations}
%

From the perspective of beamformer design, an important observation is that $\mu_{f_k}$ is a non-convex quadratic function of the beamformers $\{\mathbf{w}_k\}_{k=1}^K$, but for fixed beamforming directions $\{\mathbf{u}_k\}_{k=1}^K$ it is a linear function of the power loading $\{\beta_k\}_{k=1}^K$. The variance $\sigma_{f_k}^2$ is a quartic function of the beamformers, and for fixed directions is a non-convex quadratic function of the power loading. In scenarios in which the model $\mathbf{e}_k \backsim \mathcal{CN} (0, \sigma_{e_k}^2\mathbf{I})$ is appropriate, these expressions simplify to
\begin{subequations}
\begin{align}
  \mu_{f_k} & = \mathbf{h}_{e_k}^H \mathbf{Q}_k \mathbf{h}_{e_k}- \sigma_k^2  + \sigma_{e_k}^2 \Bigl(\beta_k /\gamma_k - \sum_{j \neq k}\beta_j \Bigr). \label{mean_eqn} \\
  \sigma_{f_k}^2 & =2 \sigma_{e_k}^2 \mathbf{h}_{e_k}^H  \mathbf{Q}_{k}^2 \mathbf{h}_{e_k} +\sigma_{e_k}^4 \text{tr} (\mathbf{Q}_{k}^2). \label{var_rel}
    \end{align}
\end{subequations}
We will focus on this simplified case in the following sections.

\section{Offset-based robust beamforming}\label{mod_dir}

As discussed above, the robust beamforming problem in \eqref{outage_min} is fundamentally hard to solve due to the intractability of the probabilistic SINR outage constraint in \eqref{sinr5}. If we were to replace that constraint with its offset  approximation, $\mu_{f_k} \geq r_k \sigma_{f_k}$, then the problem in \eqref{outage_min} can be closely approximated by
\begin{subequations}\label{outage_min2}
\begin{align}
    \min_{\substack{\mathbf{w}_k}}  \quad &  \textstyle\sum_{k=1}^K \mathbf{w}_k^H \mathbf{w}_k \\
   \text{subject to} \quad  & \mu_{f_k} \geq r_k \sigma_{f_k}, \quad \forall k. \label{sinr3}
    \end{align}
\end{subequations}
Some insights into the behaviour of solutions to  \eqref{outage_min2} can be obtained by observing that when the values of $r_k$ are chosen to be the same, the beamforming vectors are designed so that users with a large SINR variance are provided with a larger SINR mean. To do so, those users with a lower SINR variance are not provided with as large mean SINR  as they do not need the same protection against the uncertainty.

To develop an algorithm to obtain good solutions to  \eqref{outage_min2}, we observe that in \eqref{sinr3} we have the  term $\mu_{f_k}$ which is quadratic in $\mathbf{w}_k$, and we also have the term $ \sigma_{f_k}^2= 2 \sigma_{e_k}^2 \mathbf{h}_{e_k}^H  \mathbf{Q}_k^2 \mathbf{h}_{e_k} +\sigma_{e_k}^4 \text{tr} (\mathbf{Q}_k^2) $, which includes the square of the matrix $ \mathbf{Q}_k$  and, accordingly, is quartic in  $\mathbf{w}_k$. If we make the substitution $\mathbf{W}_k =\mathbf{w}_k \mathbf{w}_k^H$, then the functions in \eqref{sinr3} become linear and quadratic functions of $\mathbf{W}_k$ and the objective becomes linear. As such, the remaining difficulty in the reformulation of the problem is the set of rank-one constraints on $\mathbf{W}_k$. If we relax those constraints we obtain the following semidefinite relaxation of the problem in \eqref{outage_min2}
\begin{subequations}\label{outage_min4}
\begin{align}
    \min_{\substack{\mathbf{W}_k, d_{1k}, d_{2k}}}   \quad &\text{tr} \Bigl(\textstyle\sum_{k=1}^K  \mathbf{W}_k \Bigr) \\
   \text{s.t.} \quad  & \mathbf{h}_{e_k}^H \mathbf{Q}_k \mathbf{h}_{e_k}- \sigma_k^2  + \sigma_{e_k}^2 \text{tr}(\mathbf{W}_k) /\gamma_k   \nonumber \\
   & \quad - \sigma_{e_k}^2 \text{tr} \Bigl(\textstyle\sum_{j\neq k}  \mathbf{W}_j \Bigr)  \geq  r_k \| [d_{1k} \; d_{2k} ]\| ,  \\
   & d_{1k}  \geq   \sqrt{2} \sigma_{e_k}  \| \mathbf{h}_{e_k}^H  \mathbf{Q}_k \|, \\
    & d_{2k}  \geq \sigma_{e_k}^2 \|   \mathbf{Q}_k \|_F, \\
   &  \mathbf{W}_k \succeq \mathbf{0}, \quad \forall k,
    \end{align}
\end{subequations}
where  $\|   \cdot \|_F$ represents the Frobenius norm of the matrix.
In this formulation, each SINR constraint in \eqref{sinr3} is replaced by three second order cone (SOC) constraints. Thus, the problem in \eqref{outage_min4} is a convex conic optimization problem and can be efficiently solved using interior point methods. Two refined implementations of those methods are easily accessible through the $\textsc{Matlab}$-based CVX tool  \cite{cvx}. In our numerical experience, the rank of the optimal $\mathbf{W}_k$'s in  \eqref{outage_min4} has always been one. When that occurs, the semidefinite relaxation is tight and the optimal beamformer vectors $\mathbf{w}_k$  can be directly obtained from the optimal matrices $\mathbf{W}_k$. This phenomenon has been established in some related beamforming problems \cite{Reference2,RobustSINR,UnravelingtheRankOne}, and has been observed numerically in a number of other downlink beamforming problems; e.g., \cite{OutageConstrained}.

\subsection{Low-complexity precoding algorithm}

Although the problem in \eqref{outage_min4} is convex, it contains $3K$ SOC constraints, plus the $K$ semidefinite constraints. As a result, solving \eqref{outage_min4} incurs a significant computational load even for a moderate number of antennas. In this section, we will first show how a mild approximation of the problem in \eqref{outage_min4} leads to an optimization problem with only $K$ SOC constraints. We will then use insights from the KKT conditions of that problem to show that it can be approximately solved using the iterative evaluation of a sequence of closed-form expressions.

The approximation is based on the observation, made above, that in practical downlink systems the uncertainty in the channel estimates must be small in order for the system to support reasonable rates \cite{MIMObroadcast}. In such scenarios, the term in \eqref{sinr3} containing $\sigma_{e_k}^4$ will typically be significantly smaller than the other term. Accordingly, $\sigma_{f_k}^2 \approx 2 \sigma_{e_k}^2 \mathbf{h}_{e_k}^H  \mathbf{Q}_k^2 \mathbf{h}_{e_k}$ is a reasonable approximation.
Applying this approximation in the context of the problem in \eqref{outage_min2} we obtain the following approximation of  \eqref{sinr3}
\begin{multline}\label{new_sinr_const}
\mathbf{h}_{e_k}^H \mathbf{Q}_k \mathbf{h}_{e_k}- \sigma_k^2  + \sigma_{e_k}^2 \mathbf{w}_k^H \mathbf{w}_k /\gamma_k - \sigma_{e_k}^2 \sum_{j \neq k} \mathbf{w}_j^H \mathbf{w}_j \\ \geq  r_k  \sqrt{2} \sigma_{e_k} \|\mathbf{h}_{e_k}^H  \mathbf{Q}_k\|.
\end{multline}

The semidefinite relaxation of the resulting approximation of the problem in \eqref{outage_min2} can be written as
\begin{subequations}\label{outage_min5}
\begin{align}
    \min_{\substack{\mathbf{W}_k}, \mathbf{d}_k}   \quad &\text{tr} \Bigl(\textstyle\sum_{k=1}^K  \mathbf{W}_k \Bigr) \\
   \text{s.t.} \quad  & \mathbf{h}_{e_k}^H \mathbf{Q}_k \mathbf{h}_{e_k}- \sigma_k^2  + \sigma_{e_k}^2 \text{tr}(\mathbf{W}_k) /\gamma_k   \nonumber    \\
   & \quad - \sigma_{e_k}^2 \text{tr} \Bigl(\textstyle\sum_{j\neq k}  \mathbf{W}_k \Bigr) \geq  \| \mathbf{d}_k \|,  \label{sinr4}   \\
    & \mathbf{d}_k =  r _k \sqrt{2} \sigma_{e_k}   \mathbf{Q}_k \mathbf{h}_{e_k}, \label{dconst} \\
   &  \mathbf{W}_k \succeq \mathbf{0}, \quad \forall i.
    \end{align}
\end{subequations}
We note that the problem in \eqref{outage_min5} is over parameterized (the vectors $ \mathbf{d}_k$ are not needed), but this  over parameterization will simplify the following analysis.

The problem in \eqref{outage_min5} is another convex conic program, but it  has significantly fewer constraints than that in  \eqref{outage_min4}; there are $K$ SOC constraints rather than the $3K$ in \eqref{outage_min4}. While it can be solved with less computational effort than \eqref{outage_min4}, the presence of the semidefinite constraints means that considerable effort is still required.  To derive a more efficient algorithm, we  examine the Lagrangian of \eqref{outage_min5}, assuming that the matrices $\mathbf{W}_k$ are of rank one. If we let $\nu_k$ denote the dual variable for the constraint in \eqref{sinr4}, and $\boldsymbol\psi_{f_k}$ denote the vector of  dual variables for the equality constraint in \eqref{dconst}, the Lagrangian can be written as

\begin{multline}
     \mathcal{L}(\mathbf{w}_k, \mathbf{d}_k, \nu_k,\boldsymbol\psi_{f_k})= \sum_{k=1}^{K} \mathbf{w}_k^H \mathbf{w}_k -\sum_{k=1}^{K}\nu_k \Bigl(\mathbf{h}_{e_k}^H \mathbf{Q}_k \mathbf{h}_{e_k} - \sigma_k^2  \\ +\sigma_{e_k}^2 \mathbf{w}_k^H \mathbf{w}_k /\gamma_k - \sigma_{e_k}^2 \sum_{j \neq k}\mathbf{w}_j^H \mathbf{w}_j - \| \mathbf{d}_k \| \Bigr)     \\ -\sum_{k=1}^{K} \boldsymbol\psi_{f_k}^H (\mathbf{d}_k -  r_k  \sqrt{2} \sigma_{e_k}   \mathbf{Q}_k \mathbf{h}_{e_k}).
\end{multline}
From the KKT conditions of the problem in \eqref{outage_min5}, we can deduce that
\begin{multline}\label{closed_form}
\mathbf{w}_k =\Biggl( \frac{\nu_k}{\gamma_k}\mathbf{h}_{e_k} \mathbf{h}_{e_k}^H-\sum_{j\neq k} \nu_j \mathbf{h}_{e_j} \mathbf{h}_{e_j}^H  + \frac{\nu_k  \sigma_{e_k}^2}{\gamma_k} \mathbf{I}  -\sum_{j\neq k} \nu_j  \sigma_{e_k}^2 \mathbf{I}  \\ -  \frac{ r_k  \sqrt{2} \sigma_{e_k} }{\gamma_k}   \text{Re} \{\boldsymbol\psi_{f_k} \mathbf{h}_{e_k}^H \}  +   \sum_{j\neq k}   r_j  \sqrt{2} \sigma_{e_k}   \text{Re} \{\boldsymbol\psi_j \mathbf{h}_{e_j}^H \}    \Biggr)\mathbf{w}_k,
\end{multline}
which is an  eigen equation for the direction $\mathbf{u}_k$. Using a similar approach to the perfect CSI case \cite{OptimalMultiuserTransmit}, we can rearrange this equation to obtain the following fixed-point equation for $\nu_k$,
\begin{multline}\label{nu_mod2}
\nu_k^{-1}  = \mathbf{h}_{e_k}^H \Biggl( \mathbf{I} + \sum_j \nu_j \mathbf{h}_{e_j} \mathbf{h}_{e_j}^H  - \frac{\nu_k  \sigma_{e_k}^2}{\gamma_k} \mathbf{I}  +\sum_{j\neq k} \nu_j  \sigma_{e_k}^2 \mathbf{I}    \\  +\frac{ r_k  \sqrt{2} \sigma_{e_k} }{\gamma_k}   \text{Re} \{\boldsymbol\psi_{f_k} \mathbf{h}_{e_k}^H \}  -   \sum_{j\neq k}   r_j  \sqrt{2} \sigma_{e_k}   \text{Re} \{\boldsymbol\psi_j \mathbf{h}_{e_j}^H \}    \Biggr)^{-1} \\ \times \mathbf{h}_{e_k}  \Bigl(1+\frac{1}{\gamma_k} \Bigr).
\end{multline}
The expressions in \eqref{closed_form} and \eqref{nu_mod2} share a similar structure to those obtained for the corresponding QoS problem in the case of perfect CSI at the BS \cite{OptimalMultiuserTransmit}, but the matrix components of each equation contain four additional terms that are dependent on the variance of the channel estimation error. To exploit this structure and obtain an efficient algorithm for good solutions to \eqref{outage_min5} we observe that if we were given $\{\boldsymbol\psi_{f_k}\}$, then we could solve the fixed-point equations in \eqref{nu_mod2} for $\{ \nu_k \}$, and then we could solve the eigen equations in \eqref{closed_form} for the beamforming directions $\{\mathbf{u}_k\}$. The solution could then be completed by performing the appropriate power loading, which will be explained in the following section. Therefore, if we could find a reasonable approximation for the vectors $\boldsymbol\psi_{f_k}$, we would obtain an iterative closed-form solution. To do so, we observe that the variable $\mathbf{d}_k$ in \eqref{dconst} appears in the Lagrangian in the term $ \nu_k \| \mathbf{d}_k \| -\boldsymbol\psi_{f_k}^H \mathbf{d}_k$.
Accordingly, from the stationarity component of the KKT conditions  we have that $\| \boldsymbol\psi_{f_k}  \| =   \nu_k$ and that $\mathbf{d}_k$ and $\boldsymbol\psi_{f_k}$ are in the same direction; i.e.,  $\mathbf{d}_k/ \| \mathbf{d}_k\|   = \boldsymbol\psi_{f_k}  /\| \boldsymbol\psi_{f_k}  \|$. Accordingly, we can write
\begin{equation}\label{psi}
\boldsymbol\psi_{f_k}= \nu_k \mathbf{d}_k/ \| \mathbf{d}_k\|.
\end{equation}
Since  $\mathbf{d}_k =  r _k \sqrt{2} \sigma_{e_k}   \mathbf{Q}_k \mathbf{h}_{e_k}$, $\boldsymbol\psi_{f_k}$ explicitly depends on the beamforming directions, which have not yet been determined. However, we observe that if we substitute \eqref{psi} into \eqref{nu_mod2}, the terms involving $\mathbf{d}_k$ are multiplied by the standard deviation of the error, $\sigma_{e_k}$. As we have already argued in the derivation of the approximations that lead to \eqref{outage_min4}, $\sigma_{e_k}$ will be small in effective downlink beamforming schemes, and this suggests that reasonable initial approximations of the directions should yield a good approximation of $\{ \nu_k \}$, and hence a good set of beamforming directions. We suggest the use of the zero-forcing (ZF) directions  \cite{Zeroforcingmethods} for the estimated channels, which we will denote by $\mathbf{u}_{z_k}$. When we use that initialization, the initial direction of $\mathbf{d}_k$ will be the same as $\mathbf{u}_{z_k}$, which allows us to rewrite the fixed-point equations in \eqref{nu_mod2} as
\begin{multline}\label{nu_mod}
\nu_k^{-1}  = \mathbf{h}_{e_k}^H \Biggl( \mathbf{I} + \sum_j \nu_j \mathbf{h}_{e_j} \mathbf{h}_{e_j}^H  - \frac{\nu_k  \sigma_{e_k}^2}{\gamma_k} \mathbf{I}  +\sum_{j\neq k} \nu_j  \sigma_{e_k}^2 \mathbf{I}    \\  +\frac{ r  \sqrt{2} \sigma_{e_k} \nu_k }{\gamma_k}   \text{Re} \{\mathbf{u}_{z_k} \mathbf{h}_{e_k}^H \}  -   \sum_{j\neq k}   r  \sqrt{2} \sigma_{e_k}  \nu_j  \text{Re} \{\mathbf{u}_{z_j} \mathbf{h}_{e_j}^H \}    \Biggr)^{-1} \\ \times \mathbf{h}_{e_k}  \Bigl(1+\frac{1}{\gamma_k} \Bigr).
\end{multline}

The derivations outlined above are summarized in the sequence of closed-form operations in Alg. \ref{Alg1}. While the initial approximation can be improved by using the beamformers obtained in step 4 to obtain a refined estimate of the direction of $\mathbf{d}_k$ and returning to step 2 of the algorithm, the simulation results in Section \ref{sec_sim} suggest that the one-shot approach taken in Alg. \ref{Alg1} produces a solution whose performance is quite close to that of the original offset-based design formulation in \eqref{outage_min4}. That suggests that in the scenarios that we have considered, the underlying approximations are working quite well.

 \begin{algorithm}
\caption{Iterative closed-form beamformer design}
\label{Alg1}
\begin{algorithmic}[1]
\State Find the ZF directions $\{\mathbf{u}_{z_k}\}$.
\State Find each $\nu_k$ using \eqref{nu_mod}.
\State Find each $\mathbf{u}_k$ using  the corresponding variant of \eqref{closed_form}.
\State Apply the power loading developed in Section \ref{per_user_Power_Loading_algorithm}.
\end{algorithmic}
\end{algorithm}

\subsection{Constant-offset algorithm \cite{LowComplexityRobustMISO}}\label{sec_org_offset_max}

As is apparent from the derivation in the previous section, one of the challenges that complicates the closed-form calculations is the  quartic dependence of the variances $\sigma_{f_k}^2$ on the beamforming vectors $\mathbf{w}_k$. One way in which these complications can be reduced is to modify the offset approximation in \eqref{offset_constr} so that the mean,  $\mu_{f_k}$, is constrained to be greater than a constant; i.e., the SINR constraint is replaced by    $$ \mu_{f_k} \geq r_k. $$
If we make the approximation that the channel estimation errors are small enough that the third term on the right hand side of \eqref{mean_eqn} can be neglected, the semidefinite relaxation of the resulting approximation of \eqref{outage_min2} can be written as
\begin{subequations}\label{r_prob}
\begin{align}
    \min_{\substack{\mathbf{W}_k}}   \quad &\text{tr} \Bigl(\textstyle\sum_{k=1}^K  \mathbf{W}_k \Bigr) \\
   \text{s.t.} \quad  & \mathbf{h}_{e_k}^H \mathbf{Q}_k \mathbf{h}_{e_k}- \sigma_k^2 \geq  r_k, \label{sinr6}\\
   &  \mathbf{W}_k \succeq \mathbf{0}, \quad \forall k.
    \end{align}
\end{subequations}
Interestingly, this problem arose previously in the context of a low-complexity solution to the robust beamforming design problem for FDD and TDD systems that use a zero-outage region  approach, and the semidefinite relaxation was shown to be tight \cite{LowComplexityRobustMISO}. The zero-outage region approach provides robustness by requiring that the SINR constraints hold for all channels in a neighbourhood of the estimated channel.

The iterative closed-form solution to \eqref{r_prob} has a similar structure to that in Alg. \ref{Alg1}, but given the simpler structure of the problem, the Lagrange multipliers $\boldsymbol\psi_{f_k}$ disappear, and the expressions in \eqref{closed_form} and \eqref{nu_mod2} simplify to
\begin{equation}\label{closed_form2}
\mathbf{w}_k =\Biggl( \frac{\nu_k}{\gamma_k}\mathbf{h}_{e_k} \mathbf{h}_{e_k}^H-\sum_{j\neq k} \nu_j \mathbf{h}_{e_j} \mathbf{h}_{e_j}^H \Biggr)\mathbf{w}_k,
\end{equation}
\begin{equation}\label{nu}
\nu_k^{-1}  = \mathbf{h}_{e_k}^H \Bigl(\mathbf{I}+\textstyle\sum_{j} \nu_j \mathbf{h}_{e_j} \mathbf{h}_{e_j}^H \Bigr)^{-1} \mathbf{h}_{e_k}  \Bigl(1+\frac{1}{\gamma_k} \Bigr).
\end{equation}
After obtaining the beamforming directions from \eqref{nu} and \eqref{closed_form2}, the power loading in \cite{LowComplexityRobustMISO} is performed based on the fact that the constraints in \eqref{sinr6} are satisfied with equality at optimality. (If this were not the case for constraint $k$, then the power allocated to $\mathbf{w}_k$ could be reduced in a way that will still satisfy all the constraints and provide a lower objective value, contradicting the presumed optimality.) While doing so generates a solution to \eqref{r_prob}, significant performance gains can be obtained when the beamforming directions obtained from \eqref{closed_form2} are combined with the power loading algorithm presented in Section \ref{per_user_Power_Loading_algorithm}.

\subsection{Complexity analysis and further approximations}\label{sec_approxs}

The problems in \eqref{outage_min4} and \eqref{outage_min5} are convex optimization problems with SOC and semidefinite constraints. General purpose interior point methods for such problems require $\mathcal{O}(N_t^6)$ per iteration, which represents a significant computational load.
In contrast, the key computational steps in the iterative closed-form approximation, Alg. \ref{Alg1}, are those in  \eqref{closed_form}, \eqref{nu_mod} and the calculation of the ZF directions that are used in the initialization. The ZF directions can be obtained in  $\mathcal{O}(N_t^2 K)$ operations.
The computational cost of solving  \eqref{nu_mod} is dominated by the matrix inversion required for each user and hence it grows as $\mathcal{O}(N_t^3 K)$. We can exploit the factorized matrix structure in \eqref{closed_form} which allows for an efficient use of the power iteration method. Therefore, the cost of step 3 grows as $\mathcal{O}(N_t K^2)$. We can see that it is the computation of the Lagrange multipliers \eqref{nu_mod} that requires most of the resources to compute the beamforming directions.

The constant-offset algorithm \cite{LowComplexityRobustMISO} that was reviewed in Section~\ref{sec_org_offset_max} does not require an initial set of directions and the expression for $\nu_k$ is significantly simpler. In particular, the matrix to be inverted is the same for each user, which reduces the number of computations required to $\mathcal{O}(N_t^3)$. Furthermore, additional approximations can be applied to avoid the matrix inversion all together.
When the channels are nearly orthogonal, as they tend to be in massive MISO channels that ``harden" as the number of antennas increases \cite{Multipleantennachannelhardening}, then if we let  $\alpha_k= \|\mathbf{h}_{e_k}\|^2$, we can write $\textstyle\sum_{j} \nu_j \mathbf{h}_{e_j} \mathbf{h}_{e_j}^H$ in the form of an eigen decomposition $\textstyle\sum_{j} \nu_j  \alpha_j \frac{\mathbf{h}_{e_j}}{\sqrt{\alpha_j} }  \frac{\mathbf{h}_{e_j}^H}{\sqrt{\alpha_j}}$,
and hence,
$$\mathbf{h}_{e_k}^H \Bigl(\mathbf{I}+\textstyle\sum_{j} \nu_j \alpha_j \frac{\mathbf{h}_{e_j}}{\sqrt{\alpha_j} } \frac{\mathbf{h}_{e_j}^H}{\sqrt{\alpha_j}} \Bigr)^{-1} \mathbf{h}_{e_k} \approx \frac{\alpha_k}{1+\nu_k \alpha_k}.$$
Accordingly, we can approximate \eqref{nu} by  $$\nu_k\approx \gamma_k/\alpha_k.$$
To find the channel norms $\alpha_k=  \| \mathbf{h}_{e_k} \|^2$ we need only $\mathcal{O}(N_t)$ operations. Hence, that approximation enables us to compute all $\nu_k$s in only $\mathcal{O}(N_t K)$ operations.

\section{Offset-based robust power Loading}\label{per_user_Power_Loading_algorithm}

In this section, we will show how to apply the offset-based approach to the power loading problem that remains if the beamforming directions are chosen separately. Examples of choices for those directions include the maximum ratio transmission (MRT), zero-forcing (ZF), or regularized zero-forcing (RZF) directions, which are calculated from the estimated channels, or any of the directions generated by the previously described algorithms. Once the directions are chosen, we can  rewrite the problem in \eqref{outage_min2} as
\begin{subequations}\label{outage_min6}
\begin{align}
    \min_{\substack{\beta_k}} \     \quad & \sum_{k=1}^K \beta_k  \\
   \text{subject to} \quad  & \mu_{f_k} \geq r_k \sigma_{f_k}, \quad \forall k, \label{sinr2}
     \end{align}
\end{subequations}
where for fixed directions $\{\mathbf{u}_k\}$ the expressions for $\mu_{f_k}$ and $\sigma_{f_k}$ in \eqref{mean_eqn} and \eqref{var_rel} simplify to
\begin{subequations}\label{mean_eqn2}
\begin{align}
  \mu_{f_k} & = |\mathbf{h}_{e_k}^H \mathbf{u}_k|^2  \beta_k/\gamma_k - \sum_{j \neq k} |\mathbf{h}_{e_k}^H \mathbf{u}_j|^2  \beta_j- \sigma_k^2   \nonumber\\
  & \qquad +\sigma_{e_k}^2 \Bigl(\beta_k /\gamma_k - \sum_{j \neq k}\beta_j \Bigr).  \\
  \sigma_{f_k}^2 & =2 \sigma_{e_k}^2 \mathbf{h}_{e_k}^H   \Bigl(\beta_k \mathbf{u}_k \mathbf{u}_k^H/\gamma_k-\sum_{j \neq k}\beta_j \mathbf{u}_j \mathbf{u}_j^H \Bigr)^2 \mathbf{h}_{e_k}  \nonumber\\ &\qquad +\sigma_{e_k}^4 \text{tr} \Bigl(\beta_k \mathbf{u}_k \mathbf{u}_k^H/\gamma_k-\sum_{j \neq k}\beta_j \mathbf{u}_j \mathbf{u}_j^H \Bigr)^2. \label{var_rel2}
    \end{align}
\end{subequations}

Since $\mu_{f_k} $ is linear in $\{\beta_k\}$ and $\sigma_{f_k}$ is a convex quadratic function of $\{\beta_k\}$, the problem in  \eqref{outage_min6} can be rewritten as an SOC programming problem, and an optimal solution can be efficiently obtained using generic interior-point methods. However, to begin to develop a more efficient algorithm that exploits some of the specific features of the problem in \eqref{outage_min6}, we observe that at optimality the constraints in \eqref{sinr2} hold with equality. If this were not the case for constraint $k$, then $\beta_k $ could be reduced in a way that still satisfies the constraints and yet provides a lower objective value, which would contradict the presumed optimality. To use that observation, we note that if the variances $\sigma_{f_k}^2$ are fixed, then the set of equations $\{\mu_{k} =r_k \sigma_{f_k}\}$ yields $K$ linear equations in the $K$ design variables $\{\beta_k \}_{k=1}^K$.  If we  define $\boldsymbol{\beta}=[\beta_1, \beta_2,..., \beta_K]^T$, $\boldsymbol{\sigma}_f=[\sigma_{f_1}, \sigma_{f_2},..., \sigma_{f_K}]^T$, $\boldsymbol{\sigma}=[\sigma_{1}, \sigma_{2},..., \sigma_{K}]^T$, $\mathbf{r}=[r_1, r_2,...,r_k]^T$, and the matrix $\mathbf{A}$ such that $\mathbf{[A]}_{ii}= | \mathbf{h}_{e_i}^H {\mathbf{u}}_i |^2/\gamma_i+\sigma_{e_i}^2 /\gamma_i $, and $\mathbf{[A]}_{ij}= - | \mathbf{h}_{e_i}^H {\mathbf{u}}_j |^2 - \sigma_{e_i}^2$,  $\forall i \neq j$, then  the set of linear equations can be written as
\begin{equation}\label{A_eqn}
   \mathbf{A} \boldsymbol{\beta} =\boldsymbol{\sigma}^2+ \boldsymbol\sigma_{f} \odot \mathbf{r},
\end{equation}
in which $\odot$ represents element-by-element multiplication. Once the values of $\{\beta_k \}$ have been found, we can update the value of $\boldsymbol{\sigma}_f$ using \eqref{var_rel2}. That suggests the iterative linearization  algorithm for solving \eqref{outage_min6} that is summarized in Alg.~\ref{Alg2}.

 \begin{algorithm}
\caption{The power loading algorithm}
\label{Alg2}
\begin{algorithmic}[1]
\State Initialize $\sigma_{f_k}=1$. Compute $\mathbf{A}$ and $\mathbf{A}^{-1}$.
\State Find $\boldsymbol{\beta}$ by solving the set of linear equations in \eqref{A_eqn}.
\State Update each $\sigma_{f_k}$ using \eqref{var_rel2}.
\State Return to 2 until a termination criterion is satisfied.
\end{algorithmic}
\end{algorithm}

By observing the dependence of $\boldsymbol{\sigma}_f$ on $\boldsymbol{\beta}$ in \eqref{var_rel2}, Alg. \ref{Alg2} can be written in the form of a fixed point technique by writing $\boldsymbol{\beta} =\mathbf{A}^{-1}  \boldsymbol{\sigma}^2+ \mathbf{A}^{-1} (\boldsymbol\sigma_{f}  \odot \mathbf{r})$. The eigenvalues of $\mathbf{A}^{-1}$ determine the convergence properties for these fixed-point equations. Since the matrix $\mathbf{A}$ typically has large diagonal values representing the signal powers, and lower values on the off-diagonal elements representing the interference powers, the eigen values of $\mathbf{A}^{-1}$ will typically be less than one. Our numerical experience not only confirms this observation, but also suggests that the number of iterations needed for near-optimal performance is very small. In terms of computational cost, the initialization step in Alg. \ref{Alg2} requires $\mathcal{O}(K^2 N_t)$ operations to compute $\mathbf{A}$ and  $\mathcal{O}(K^3)$ operations to compute $\mathbf{A}^{-1}$. In each iteration the computational cost for step 2 is  $\mathcal{O}(K^2)$ operations, and the cost of step 3 is $\mathcal{O}(K N_t^2)$  operations.

\subsection{Simplifying the SINR variance calculation}\label{simplified_var_subsect}

The above analysis shows that the only step in Alg. \ref{Alg2} whose computational cost grows faster than linearly in the number of antennas is the computation of  $\sigma_{f_k}$. In massive MISO systems, the resulting  computational load can be significant. To reduce the required computations, we observe that when the number of antennas is large and the channels are uncorrelated, the inner product between different channels will typically  be relatively small. Since the beamforming directions will typically be closely aligned with the channel vectors, the inner product between different beamforming vectors will likely be small as well. This observation suggests removing the cross terms $\mathbf{u}_j^H \mathbf{u}_k, \forall j \neq k$ in \eqref{var_rel2}. That would yield the following approximations
\begin{equation}
\begin{aligned}
     \mathbf{h}_{e_k}^H  \mathbf{Q}_{k} \mathbf{Q}_{k} \mathbf{h}_{e_k} &=   \mathbf{h}_{e_k}^H \Bigl(\beta_k \mathbf{u}_k \mathbf{u}_k^H/\gamma_k-\sum_{j \neq k}\beta_j \mathbf{u}_j \mathbf{u}_j^H \Bigr)^2  \mathbf{h}_{e_k} \\
               &\approx    |\mathbf{h}_{e_k}^H \mathbf{u}_k|^2  \beta_k^2/\gamma_k^2 + \sum_{j \neq k} |\mathbf{h}_{e_k}^H \mathbf{u}_j|^2  \beta_j^2,
   \end{aligned}
\end{equation}
and
\begin{equation}
\begin{aligned}
     \text{tr} (\mathbf{Q}_{k}^2)&=   \text{tr} \Bigl(\beta_k \mathbf{u}_k \mathbf{u}_k^H/\gamma_k-\sum_{j \neq k}\beta_j \mathbf{u}_j \mathbf{u}_j^H \Bigr)^2 \\
     &\approx   \text{tr} \Bigl(\beta_k^2 \mathbf{u}_k \mathbf{u}_k^H \mathbf{u}_k \mathbf{u}_k^H /\gamma_k^2+\sum_{j \neq k}\beta_j^2 \mathbf{u}_j \mathbf{u}_j^H \mathbf{u}_j \mathbf{u}_j^H \Bigr) \\
     &= \beta_k^2/\gamma_k^2+\sum_{j \neq k}\beta_j^2.
   \end{aligned}
\end{equation}
The numerical results presented in Section~\ref{sec_sim} indicate that these approximations result in designs that are  very close in performance to those obtained from the original formulations, even when the number of antennas is quite small. Furthermore, since the terms $|\mathbf{h}_{e_k}^H \mathbf{u}_j|^2$ are already computed in the initialization step that constructs the matrix $\mathbf{A}$, these approximations reduce the computational cost of updating $\boldsymbol\sigma_f$ in step 3 of Alg.~\ref{Alg2} from $\mathcal{O}(N_t^2 K)$ to $\mathcal{O}(K^2)$.

\subsection{User rescheduling}\label{userrescheduling}

One of the fundamental characteristics of the original outage constrained beamformer design problem in \eqref{outage_min2} is that for a certain set of channel estimates the problem may be infeasible. That is, there may be no set of beamformers that can satisfy the outage constraints. Furthermore, even when the problem is feasible, the solution may be impractical in the sense that the minimum transmission power required to satisfy the outage constraints may exceed the capability of the BS. The approximations of the original formulation in \eqref{outage_min4} and \eqref{outage_min5} retain these characteristics, and the power loading problem in  \eqref{outage_min6} retains them, too. Fortunately, as we now explain, for systems in which each user specifies the same value for $r$, the structure of a closely-related power loading problem provides insights into which users should be rescheduled in order for the problem in \eqref{outage_min6} to be feasible, and for the solution of the problem to be within the capabilities of the BS. The auxiliary power loading problem that we will consider is that of maximizing a common offset coefficient subject to an explicit power constraint, namely
\begin{subequations}\label{outage_min7}
\begin{align}
    \max_{\substack{\beta_k},r}   \quad & r \\
    \text{subject to} \quad  &   \textstyle\sum_{k=1}^K \beta_k \leq P_t, \label{pwr_cont} \\
   \quad  & \mu_{k} \geq r \sigma_{f_k}, \quad \forall k \label{sinr7},
    \end{align}
\end{subequations}
where $P_t$ denotes the maximum transmission power of the BS. This problem is always feasible whenever all the estimated channels are different. (The value of $r$ can be decreased until all components of \eqref{sinr7} can be satisfied using a power loading that satisfies \eqref{pwr_cont}.) However, negative values and small positive values of $r$  correspond to cases with high probability of outage. The problem in \eqref{outage_min7} can be solved using an algorithm similar to that in Alg.~\ref{Alg2}. However, at the step analogous to step 2 of  Alg.~\ref{Alg2}, we need an additional equation to determine the value for $r$. That equation arises from observing that the power constraint in \eqref{pwr_cont} holds with equality at optimality, and hence, from \eqref{A_eqn} and  \eqref{pwr_cont} we have that
$$ r = \frac{P_t- \boldsymbol{1}^T \mathbf{A}^{-1} \boldsymbol\sigma^2}{ \boldsymbol{1}^T\mathbf{A}^{-1} \boldsymbol\sigma_{f}},$$
where $\boldsymbol{1}$ is the vector with all elements equal to one. This equation clearly demonstrates the relationship between the power budget and the robustness. More importantly, it shows that the users that correspond to the largest elements of  $\mathbf{A}^{-1} \boldsymbol\sigma^2$  are the ones that play the biggest role in constraining the extent of robustness that can be obtained. That suggests that if the optimal value of $r$ in  \eqref{outage_min7} is not large enough to provide the desired robustness level, one or more of those users corresponding to large values of $\mathbf{A}^{-1} \boldsymbol\sigma^2$ should be rescheduled. (We note that the use of good user selection algorithms, e.g., \cite{Ontheoptimalityofmultiantenna}, prior to the design of the beamforming directions will reduce the need to reschedule users, but the inherent capability of the proposed power loading algorithms to perform rescheduling provides significant performance gains when the initial user selection is imperfect.)

Once the optimal value of the auxiliary problem in \eqref{outage_min7} exceeds the desired value for $r$, the power minimization problem in \eqref{outage_min6} can be solved. Since the distribution of $f_k(\mathbf{e}_{k})$ is dominated by the Gaussian terms, values of $r$ in the range of 2 to 5 would be sufficient to obtain outage probabilities consistent with the expectations of contemporary applications; see Appendix~\ref{r_value_sel}.

\subsection{Average outage}\label{mod_algo}

The design formulations that we have considered up until this point have taken the form of minimization of the transmission power subject to (an approximation of) an outage constraint on each user for the current realizations of the channels. However, as we now illustrate, the proposed design approach is quite flexible and can accommodate other notions of outage.

\begin{figure}
\begin{center}
    \epsfysize= 2.4in
     \epsffile{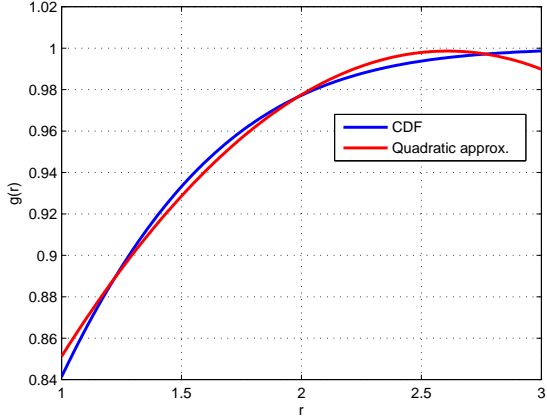}
\caption{The CDF of the standardized normal distribution, $\mathcal{N} (0, 1)$, denoted $g(r)$, and its least squares quadratic approximation over $r \in[ 1,3]$.
}\label{fig1}
\end{center}
\end{figure}

Let us assume that we have the optimal power loading and offset coefficient for the problem in \eqref{outage_min7}, which provide all the users with essentially the same outage probability. We will denote those values by  $\{\beta_k^\star\}_{k=1}^K$ and $r^\star$. Given this solution, the goal of this section is to perturb the value of the offset coefficient for each user so as to minimize the average outage probability over the users, and to adjust the power loading accordingly.  To do so, we let $\delta_{r_k}$ denote the perturbation on the $k$th user's offset coefficient; i.e., $r_k=r^\star+\delta_{r_k}$.
As discussed in Section~\ref{sect3}, in typical operating scenarios the distribution of $f_k(\mathbf{e}_k)$ can be accurately approximated by a Gaussian distribution. In that case, the outage probability for a given value of the offset coefficient $r_k$ is simply the value of the complementary cumulative distribution function (CCDF) of the standardized normal distribution, $\mathcal{N} (0, 1)$, at the value of $r_k$. If we let  $g(\cdot)$  denote the CDF of the standard normal distribution, the the problem of minimizing the outage probability becomes
\begin{subequations}
\begin{align}
    \max_{\substack{\beta_k, \delta_{r_k}}} \quad  &  \textstyle\sum_{k=1}^K g(r^\star+\delta_{r_k}) \\
   \text{s.t.} \quad  & \textstyle\sum_{k=1}^K \beta_k = P_t,
    \end{align}
\end{subequations}
where the condition $\textstyle\sum_{k=1}^K \beta_k = P_t$ ensures that the power used after perturbation will be the same as that used by the solution to \eqref{outage_min7}. That constraint can be shown to be equivalent to the linear constraint $\boldsymbol{1}^T \mathbf{A}^{-1} (\boldsymbol\sigma_{s} \odot \boldsymbol{\delta_{r}} )=0$, where $\boldsymbol{\delta_{r}}$ is the vector containing the scalars $\delta_{r_k}$. Furthermore, the CDF  $g(\cdot)$ can be well approximated by a quadratic curve; see Fig.~\ref{fig1}.
With this approximation in place, the problem can be stated as the following convex problem in $\boldsymbol\delta_{r}$
\begin{subequations}\label{pert_eqn}
\begin{align}
    \max_{\substack{\boldsymbol{\delta_{r}}}}  \quad  &  \textstyle\sum_{k=1}^K a_0(r^\star+\delta_{r_k})^2+ a_1 (r^\star+\delta_{r_k}) +a_2\\
   \text{s.t.} \quad  & \boldsymbol{1}^T \mathbf{A}^{-1} (\boldsymbol\sigma_{s} \odot \boldsymbol{\delta_{r}})=0,
    \end{align}
\end{subequations}
where $a_0, a_1$ and $a_2$ are the coefficients of the quadratic approximation of $g(r)$.
If we let $\mathbf{b}= (\boldsymbol{1}^T \mathbf{A}^{-1}) \odot \boldsymbol\sigma_{s}$, then using an analysis of the KKT conditions of \eqref{pert_eqn}, we can derive the dual variable $\zeta$ of the equality constraint as
$$\zeta=\frac{-(2 a_0 r^\star +a_1) \mathbf{b}^H \boldsymbol{1}}{\mathbf{b}^H \mathbf{b}},$$
and the required $\boldsymbol{\delta_{r}}$ as
$$\boldsymbol{\delta_{r}}=\frac{-(2 a_0 r^\star + a_1)\boldsymbol{1} -\zeta  \mathbf{b}  }{2 a_0}.$$
Accordingly, whenever we have the optimal solution $\{\beta_k^\star\}_{k=1}^K$ and $r^\star$ of the problem in \eqref{outage_min7}, we can calculate $\zeta$, and the resulting perturbations of the offset coefficient $\boldsymbol{\delta_{r}}$. The modified offset coefficient vector $\mathbf{r}$ can be updated using $r_k=r^\star+\delta_{r_k}$. The power loading $\{\beta_k\}_{k=1}^K$ is then updated  by using the linear equations arising from  \eqref{sinr7} holding with equality; i.e., $\boldsymbol{\beta} =\mathbf{A}^{-1}  \boldsymbol{\sigma}^2+ \mathbf{A}^{-1} (\boldsymbol\sigma_{f}  \odot \mathbf{r})$.

\section{Simulation results}\label{sec_sim}

In this section, we will provide three sets of numerical results. First, we will  provide simulation results that show the validity of the offset-based algorithms and compare the performance of the algorithms presented here to that of zero-outage region algorithms that obtain robustness by ensuring that outage does not occur for uncertainties that lie in a given region. Specifically we will compare with the sphere bounding (SB) algorithm presented in \cite{OutageConstrained}.  Second, we will provide comparisons between the performance of the offset-based power loading algorithms proposed in Section \ref{per_user_Power_Loading_algorithm}, the optimal power loading algorithm in \cite{Coordinateupdate}, and the perturbation-based power loading algorithm that seeks to minimize the averaged outage, which was presented in Section~\ref{mod_algo}. In the third set of simulation results, we will demonstrate the performance gains that can be obtained by using the user rescheduling and the power saving described in Section~\ref{userrescheduling}. We will also show the validity of the low-complexity approximations presented in Section~\ref{sec_approxs}.

For the initial simulation setup, we will we consider a downlink system in which a BS serves three single-antenna users. We will assume that the BS has four antennas, and the three users are randomly distributed within a radius of 3.2km. The large scale fading is described by a path-loss exponent of 3.52 and log-normal shadow fading with 8dB standard deviation, and the small scale fading is modelled using the standard i.i.d. Rayleigh model. The channel estimation error is assumed to be zero-mean and Gaussian with covariance $\sigma_{e_k}^2 \mathbf{I}$. The receiver noise level is -90dBm, and the  SINR target is set to 6dB. A simple channel-strength user selection technique is employed, where users are served only if $ 100\|\mathbf{h}_{e_k}\|^2/ \sigma_k^2 \geq \gamma_k$, where we consider 100 here as the implicit total power constraint.

Each of the algorithms that we consider involves a choice of a robustness measure. For the algorithms provided in this paper the robustness measure is the value of the offset coefficient $r_k$. For the sphere bounding algorithm in \cite{OutageConstrained} it is the size of the zero-outage region, and for the power loading algorithm in \cite{Coordinateupdate} it is directly the outage probability. To plot the performance curves, we randomly generate a set of channel realizations and provide the BS with estimates of those channels. Each algorithm is then used to design a set of beamformers that should provide the specified robustness. Using those beamformers we determine whether or not any user in the system with the actual channel realizations is in outage,  and we calculate the corresponding transmission power. By repeating this experiment over thousands of channel realizations, we can plot the average outage probability over the users  versus the average transmission  power for the different algorithms when these algorithms provide a  viable solution; by which we mean a solution that satisfies the constraints using a transmitted power that is less than 100. In fairness to all methods, the average is taken over those channel realizations for which all methods produce a viable solution.

\begin{figure}
\begin{center}
    \epsfysize= 2.8in
     \epsffile{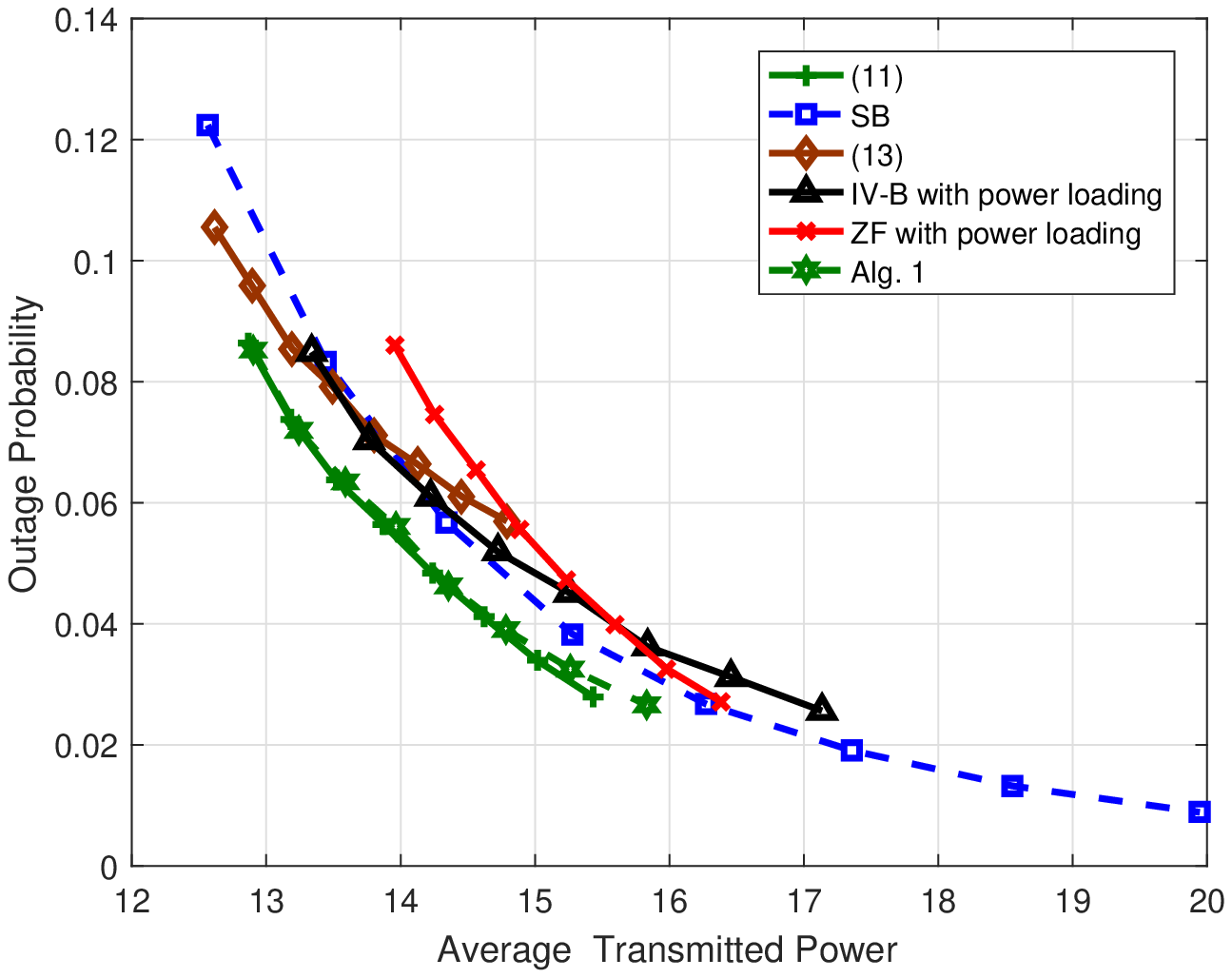}
\caption{The average transmitted power against the outage probability for a system with 3  users, 4 BS antennas, $\gamma$ = 6dB, and $\sigma_{e_k}=0.1$.
}\label{sim1}
\end{center}
\end{figure}

\begin{figure}
\begin{center}
    \epsfysize= 2.8in
     \epsffile{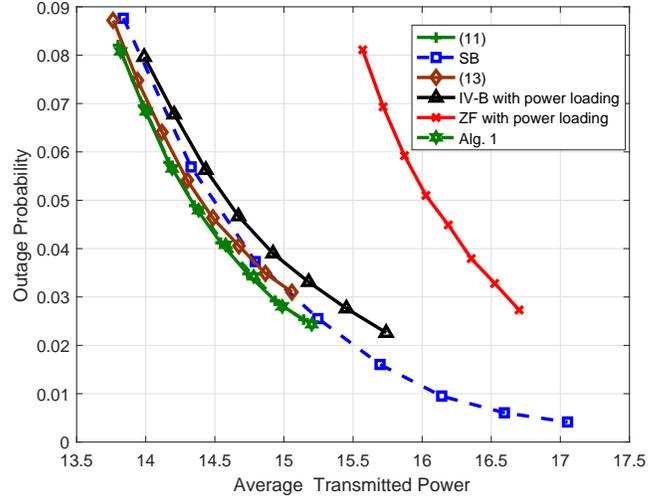}
\caption{The average transmitted power against the outage probability for a system with 3  users, 4 BS antennas, $\gamma$ = 6dB, and $\sigma_{e_k}=0.05$.
}\label{sim2}
\end{center}
\end{figure}

In Fig.~\ref{sim1}, assuming $\sigma_{e_k}=0.1$, we plot the average outage probability  versus  the average total transmitted power for the proposed robust beamforming algorithms  in \eqref{outage_min4}, \eqref{outage_min5}, Alg. \ref{Alg1}, and that of a system with the constant-offset directions described in Section.~\ref{sec_org_offset_max} and the suggested power loading in Section \ref{per_user_Power_Loading_algorithm}. As benchmarks, we plot the performance of the SB algorithm \cite{OutageConstrained}, and that of a system that employs  the ZF directions combined with the power loading in Section \ref{per_user_Power_Loading_algorithm}. In Fig.~\ref{sim2}, we repeat the experiment for $\sigma_{e_k}=0.05$. We observe that the performance gap between the proposed algorithms becomes smaller when the error variance decreases, which justifies the validity of the approximations for small error size. We also note that the performance of the low-complexity robust beamforming algorithm in Alg. \ref{Alg1} is very close to that of the original formulation in \eqref{outage_min4}, and that both algorithms provide better performance than the SB algorithm (which incurs a significantly larger computational load). The relative performance of the ZF-based algorithm with the proposed power loading algorithm in Section~\ref{per_user_Power_Loading_algorithm} depends on the uncertainty size, where comparatively better performance results are obtained when the uncertainty size is larger. That observation means that while both the offset-based beamforming directions and power loading contribute to the excellent performance for small uncertainty size, as the uncertainty size increases the role of the offset-based power loading becomes more significant. The performance of the combination of the original constant-offset directions in Section~\ref{sec_org_offset_max} with the suggested power loading in Section~\ref{per_user_Power_Loading_algorithm} is not quite as good as that of the other offset-based approaches. However, decoupling the design of the beamforming directions and that of the power loading significantly reduces the computational cost (see Section~\ref{sec_approxs}), and greatly increases the flexibility of the design, as explained in Section~\ref{per_user_Power_Loading_algorithm}.

\begin{figure}
\begin{center}
    \epsfysize= 2.8in
     \epsffile{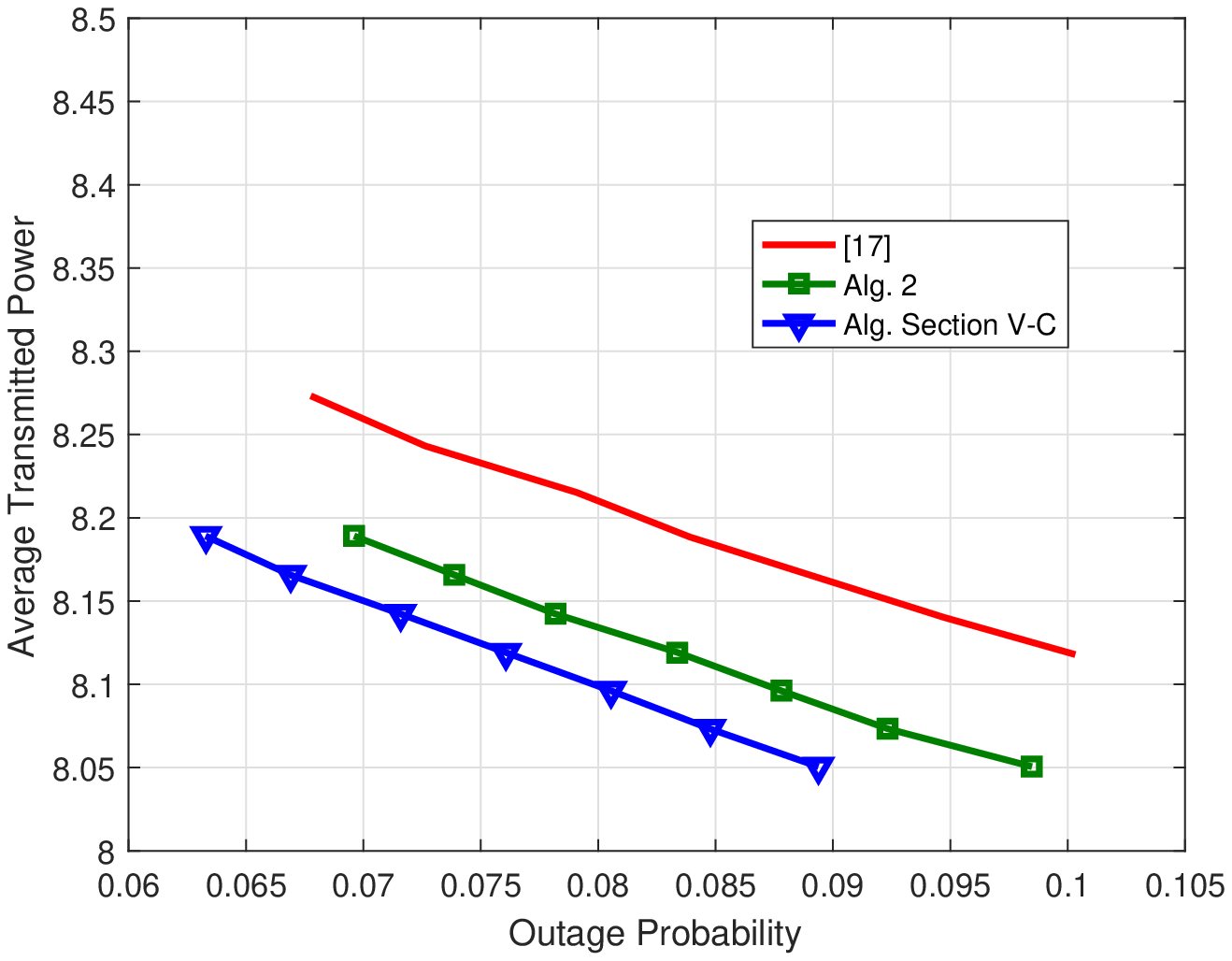}
\caption{The average transmitted power against the outage probability for a system with 3  users, 4 BS antennas, $\gamma$ = 6dB, and $\sigma_{e_k}=0.1$.
}\label{sim3}
\end{center}
\end{figure}

The second set of simulation results examines the performance gap between the proposed power loading algorithms in Section \ref{per_user_Power_Loading_algorithm}, and the power loading algorithm in \cite{Coordinateupdate} when the constant-offset directions are chosen; see Section.~\ref{sec_org_offset_max}.
In Fig.~\ref{sim3}, we plot the average outage probability versus the average transmitted power for the power loading algorithm in \cite{Coordinateupdate}, the power loading in Alg. \ref{Alg2}, and the modified power loading in Section~\ref{mod_algo}. (For the latter case, the quadratic approximation used in \eqref{pert_eqn} is the least-squares approximation in Fig.~\ref{fig1}.) While the algorithm in \cite{Coordinateupdate} is optimal in terms of the  power required to achieve the specified outage probabilities for each user and for each channel realization, the proposed algorithms provide better average outage probability. This performance is achieved while requiring no more than five iterations in the power loading algorithm in Alg. \ref{Alg2}. As one would expect, the modified power loading algorithm in Section~\ref{mod_algo} provides an even lower average outage probability than that obtained by Alg. \ref{Alg2}.

\begin{figure}
\begin{center}
    \epsfysize= 2.8in
     \epsffile{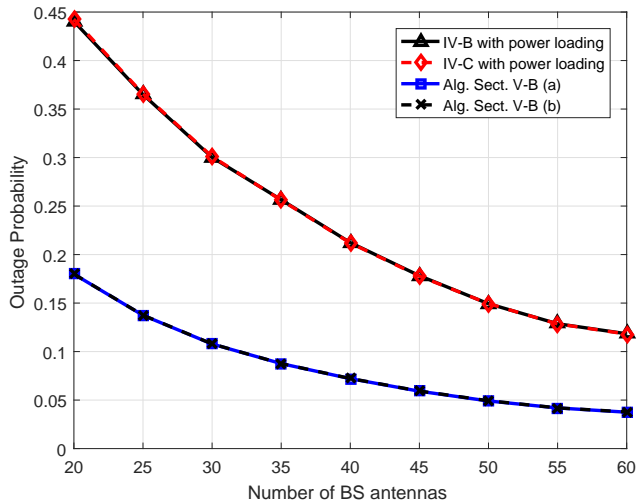}
\caption{The outage probability versus the number of antennas for a system with 6  users, $\gamma$ = 6dB, and $\sigma_{e_k}=0.1$.
}\label{sim4}
\end{center}
\end{figure}

To assess the performance gains that result from the power control capabilities of the proposed power loading algorithms, we plot the outage probability of the problem in  \eqref{outage_min7} with the constant-offset directions in Section~\ref{sec_org_offset_max} versus the number of antennas. In this case, we set the total power constraint $P_t=1$, and the number of users to six. We also plot the corresponding results when the approximations for obtaining the directions in Section~\ref{sec_approxs}, and those for obtaining the power loading in Section~\ref{simplified_var_subsect} are applied. In addition, we plot the performance of the proposed user rescheduling scheme (Alg. Sect. \ref{userrescheduling} (a)) and the user rescheduling when combined with the power saving (Alg. Sect. \ref{userrescheduling} (b)). We applied  user rescheduling whenever the resulting offset $r$ in \eqref{outage_min7} is smaller than two, and the rescheduled user(s) are considered to be in outage. For the power saving algorithm we upper bound $r$ by 5. We observe from Fig.~\ref{sim4} that the proposed approximations provide almost the same outage performance over the whole range of antenna numbers. We also observe that the user rescheduling technique greatly enhances the outage performance, especially when the number of antennas is relatively low. (When the number of antennas is low, there is a greater probability of the channels not being sufficiently orthogonal.) Fig.~\ref{sim4} shows that the power saving algorithm (Alg. Sect.~\ref{userrescheduling} (b)) provides essentially the same performance as the regular algorithms, but significant power can be saved; the average actual transmitted powers used for that  algorithm  when the number of antennas are $[20,25,\cdots,60]$ are $[0.74, 0.71, 0.67, 0.65, 0.62, 0.59, 0.56, 0.54, 0.52]$ all of which are significantly smaller than  the total power constraint $P_t=1$.

\section{Conclusion}

In this paper, a new offset-based approach is proposed for robust downlink beamforming. The approach is based on rewriting the SINR outage constraint as a non-negativity constraint on an indefinite quadratic function of the error in the base station's model of the channel. That non-negativity is then approximated by an offset-based constraint in which the mean of the function is required to be larger than a specific multiple of its standard deviation. This approach enabled the formulation of the robust beamforming design problem as a problem that can be transformed into a convex problem through the process of semidefinite relaxation (SDR). The computational complexity of the SDR problem can be further reduced when the uncertainty size is small, allowing for an iterative closed-form solution. When the beamforming directions are defined in advance, the offset-based approach generates a power loading algorithm that provides excellent performance and unique power control capabilities, while incurring  only a small computational cost. The demonstrated performance gains, and the significant differences in computational cost exemplify the advantages of using the offset-based approach instead of the sphere bounding approach. Within the suite of algorithms generated by the offset based approach, the separation of the design into the constant-offset directions (Section~\ref{sec_org_offset_max}) and the proposed power loading (Section~\ref{per_user_Power_Loading_algorithm}) provides a compelling balance between performance, computational cost and design flexibility.

%


\appendices

\section{Choice of $r_k$} \label{r_value_sel}

From Cantelli's Inequality, which is sometimes referred to as the one-sided Chebyshev inequality, we know that for any random variable $\mathbf{X}$ with mean $\mu_x$ and variance $\sigma_x$,
$$\text{Prob}(\mathbf{X}-\mu_x \leq -r \sigma_x) \leq \frac{1}{1+r^2}.$$
Therefore, if we ensure that $\mu_x \geq r \sigma_x$ then $\text{Prob}(\mathbf{X} \leq 0) \leq \frac{1}{1+r^2}$. Accordingly, if we set $r_k=\sqrt{1/\delta_k-1}$, then the approximation in \eqref{offset_constr} is ``safe", in the sense that any solution to the corresponding problem in \eqref{outage_min2} is guaranteed to satisfy the original outage constraints in \eqref{outage_min}. However, for the distributions that typically arise in downlink beamforming Cantelli's Inequality is quite loose and the resulting beamformer design is quite conservative. Indeed, as we explained in Section~\ref{sect3}, for small uncertainties the distribution of $f_k(\mathbf{e}_k)$ is close to being Gaussian. If it were in fact Gaussian, then if the beamformers are designed such that $ \mu_{f_k} \geq r_k \sigma_{f_k}$ then the outage probability would be $Q(r_k)=\frac{1}{2} \text{erfc}(\frac{r_k}{\sqrt{2}})$, where $\text{erfc}(\cdot)$ is the complementary error function.

\section{Mean and variance derivations} \label{mean_var_der}

A Gaussian random variable $\mathbf{e}_k \backsim \mathcal{CN} (\mathbf{m}_k, \mathbf{C}_k)$ can be represented as $\mathbf{e}_k =\mathbf{m}_k + \mathbf{C}_k^{1/2} \hat{\mathbf{e}}_k$, where $\hat{\mathbf{e}}_k \backsim \mathcal{CN} (0, \mathbf{I})$. Using that representation we can write

\begin{equation}
\begin{aligned}
 \mu_{f_k}& = \mathbb{E} \{f_k(\mathbf{e}_k)\} \\
&=(\mathbf{h}_{e_k}+\mathbf{m}_k)^H \mathbf{Q}_k (\mathbf{h}_{e_k}+\mathbf{m}_k) - \sigma_k^2 \\
&\quad +\mathbb{E} \{\hat{\mathbf{e}}_{k}^H \mathbf{C}_k^{1/2}  \mathbf{Q}_{k}\mathbf{C}_k^{1/2}  \hat{\mathbf{e}}_{k}\}\\
&= (\mathbf{h}_{e_k}+\mathbf{m}_k)^H \mathbf{Q}_k (\mathbf{h}_{e_k}+\mathbf{m}_k) - \sigma_k^2  + \mathbf{w}_k^H \mathbf{C}_k \mathbf{w}_k /\gamma_k  \\
 &\quad -\sum_{j \neq k} \mathbf{w}_j^H \mathbf{C}_k  \mathbf{w}_j^H.
   \end{aligned}
\end{equation}
The variance can be expressed as
\begin{equation}
\begin{aligned}
     \sigma_{f_k}^2&=\text{var}\{f_k(\mathbf{e}_{k}) \} \\
     &=\text{var}\{ 2 \text{Re}( \hat{\mathbf{e}}_{k}^H \mathbf{C}_k^{1/2} \mathbf{Q}_{k} (\mathbf{h}_{e_k}+\mathbf{m}_k  )) \\
     &\qquad  + \hat{\mathbf{e}}_{k}^H \mathbf{C}_k^{1/2}  \mathbf{Q}_{k}\mathbf{C}_k^{1/2}  \hat{\mathbf{e}}_{k} \bigr\} \\
&=2  (\mathbf{h}_{e_k}+\mathbf{m}_k)^H   \mathbf{C}_k^{1/2} \mathbf{Q}_{k}^2 \mathbf{C}_k^{1/2} (\mathbf{h}_{e_k}+\mathbf{m}_k)  \\
&\qquad +\text{var}\{\hat{\mathbf{e}}_{k}^H \mathbf{C}_k^{1/2}  \mathbf{Q}_{k}\mathbf{C}_k^{1/2}  \hat{\mathbf{e}}_{k}\}+0^* \\
&=2  (\mathbf{h}_{e_k}+\mathbf{m}_k)^H   \mathbf{C}_k^{1/2} \mathbf{Q}_{k}^2 \mathbf{C}_k^{1/2} (\mathbf{h}_{e_k}+\mathbf{m}_k) \\
&\qquad + \text{tr} (\mathbf{C}_k^{1/2} \mathbf{Q}_{k} \mathbf{C}_k^{1/2} )^2,
   \end{aligned}
\end{equation}
where $\mathbf{[A]}_{ij}$ denotes the $(i,j)$th element of the matrix $\mathbf{A}$, and tr denotes the trace function. At the point marked with the asterisk we have used the fact that the expectation of the cross terms is equal to zero. This is true because $\mathbb{E} \{2 \text{Re}( \mathbf{e}_{k}^H  \mathbf{Q}_{k} (\mathbf{h}_{e_k}+\mathbf{m}_k  )) (\hat{\mathbf{e}}_{k}^H \mathbf{C}_k^{1/2}  \mathbf{Q}_{k}\mathbf{C}_k^{1/2}  \hat{\mathbf{e}}_{k}) \}$ consists of terms containing either similar or different components from the  $\hat{\mathbf{e}}_{k} $ vector. Since $\hat{\mathbf{e}}_{k} $ has a zero mean, all terms with different indices will have a zero mean, while terms of similar indexes will take the form of a complex Gaussian raised to the power of three, which also has zero mean.


\begin{thebibliography}{50}
\providecommand{\url}[1]{#1}
\csname url@samestyle\endcsname
\providecommand{\newblock}{\relax}
\providecommand{\bibinfo}[2]{#2}
\providecommand{\BIBentrySTDinterwordspacing}{\spaceskip=0pt\relax}
\providecommand{\BIBentryALTinterwordstretchfactor}{4}
\providecommand{\BIBentryALTinterwordspacing}{\spaceskip=\fontdimen2\font plus
\BIBentryALTinterwordstretchfactor\fontdimen3\font minus
  \fontdimen4\font\relax}
\providecommand{\BIBforeignlanguage}[2]{{%
\expandafter\ifx\csname l@#1\endcsname\relax
\typeout{** WARNING: IEEEtran.bst: No hyphenation pattern has been}%
\typeout{** loaded for the language `#1'. Using the pattern for}%
\typeout{** the default language instead.}%
\else
\language=\csname l@#1\endcsname
\fi
#2}}
\providecommand{\BIBdecl}{\relax}
\BIBdecl


\bibitem{SymbollevelandMulticast}
M.~Alodeh, D.~Spano, A.~Kalantari, C.~Tsinos, D.~Christopoulos, S.~Chatzinotas,
  and B.~Ottersten, ``Symbol-level and multicast precoding for multiuser
  multiantenna downlink: A survey, classification and challenges,'' \emph{arXiv
  preprint arXiv:1703.03617}, 2017.

\bibitem{TheCapacityRegion}
H.~Weingarten, Y.~Steinberg, and S.~Shamai, ``The capacity region of the
  {G}aussian multiple-input multiple-output broadcast channel,'' \emph{{IEEE}
  Trans. Inf. Theory}, vol.~52, no.~9, pp. 3936--3964, Sep. 2006.

\bibitem{Avectorperturbationtechnique2}
B.~M. Hochwald, C.~B. Peel, and A.~L. Swindlehurst, ``A vector-perturbation
  technique for near-capacity multiantenna multiuser communication-part ii:
  perturbation,'' \emph{IEEE Trans. Commun.}, vol.~53, no.~3, pp. 537--544,
  March 2005.

\bibitem{latticereductionaided}
C.~Windpassinger, R.~F.~H. Fischer, and J.~B. Huber, ``Lattice-reduction-aided
  broadcast precoding,'' \emph{IEEE Trans. Commun.}, vol.~52, no.~12, pp.
  2057--2060, Dec. 2004.

\bibitem{Precodinginmultiantenna}
C.~Windpassinger, R.~F.~H. Fischer, T.~Vencel, and J.~B. Huber, ``Precoding in
  multiantenna and multiuser communications,'' \emph{IEEE Trans. Wireless
  Commun.}, vol.~3, no.~4, pp. 1305--1316, July 2004.

\bibitem{RobustTransmissioninDownlink}
H.~Joudeh and B.~Clerckx, ``Robust transmission in downlink multiuser {MISO}
  systems: A rate-splitting approach,'' \emph{IEEE Trans. Signal Process.},
  vol.~64, no.~23, pp. 6227--6242, Dec. 2016.

\bibitem{ConstructiveMultiuserInterference}
M.~Alodeh, S.~Chatzinotas, and B.~Ottersten, ``Constructive multiuser
  interference in symbol level precoding for the {MISO} downlink channel,''
  \emph{IEEE Trans. Signal Process.}, vol.~63, no.~9, pp. 2239--2252, May 2015.

\bibitem{ShiftingtheMIMO}
D.~Gesbert, M.~Kountouris, R.~W. Heath, C.-B. Chae, and T.~Salzer, ``Shifting
  the {MIMO} paradigm,'' \emph{IEEE Signal Proc. Mag.}, vol.~24, no.~5, pp.
  36--46, Sept. 2007.

\bibitem{Jointoptimal}
F.~Rashid-Farrokhi, L.~Tassiulas, and K.~J.~R. Liu, ``Joint optimal power
  control and beamforming in wireless networks using antenna arrays,''
  \emph{{IEEE} Trans. Commun.}, vol.~46, no.~10, pp. 1313--1324, Oct. 1998.

\bibitem{Reference2}
M.~Bengtsson and B.~Ottersten, ``Optimal and suboptimal transmit beamforming,''
  in \emph{Handbook of Antennas in Wireless Communications}, L.~C. Godara,
  Ed.\hskip 1em plus 0.5em minus 0.4em\relax CRC Press, 2001, ch.~18.

\bibitem{Solutionofthemultiuser}
M.~Schubert and H.~Boche, ``Solution of the multiuser downlink beamforming
  problem with individual {SINR} constraints,'' \emph{IEEE Trans. Veh. Tech.},
  vol.~53, no.~1, pp. 18--28, Jan. 2004.

\bibitem{OptimalMultiuserTransmit}
E.~Bjornson, M.~Bengtsson, and B.~Ottersten, ``Optimal multiuser transmit
  beamforming: A difficult problem with a simple solution structure,''
  \emph{IEEE Signal Process. Mag.}, vol.~31, no.~4, pp. 142--148, July 2014.

\bibitem{Optimalpowercontrol}
S.~Kandukuri and S.~Boyd, ``Optimal power control in interference-limited
  fading wireless channels with outage-probability specifications,'' \emph{IEEE
  Trans. Wireless Commun.}, vol.~1, no.~1, pp. 46--55, Jan. 2002.

\bibitem{Probabilisticallyconstrained}
M.~B. Shenouda and T.~N. Davidson, ``Probabilistically-constrained approaches
  to the design of the multiple antenna downlink,'' in \emph{Conf. Rec. 42nd
  Asilomar Conf. Signals, Systems, Computers}, Pacific Grove, CA, Oct. 2008,
  pp. 1120--1124.

\bibitem{OutageConstrained}
K.-Y. Wang, A.-C. So, T.-H. Chang, W.-K. Ma, and C.-Y. Chi, ``Outage
  constrained robust transmit optimization for multiuser {MISO} downlinks:
  Tractable approximations by conic optimization,'' \emph{IEEE Trans. Signal
  Process.}, vol.~62, no.~21, pp. 5690--5705, Nov. 2014.

\bibitem{LowComplexityRobustMISO}
M.~Medra, Y.~Huang, W.~K. Ma, and T.~N. Davidson, ``Low-complexity robust
  {MISO} downlink precoder design under imperfect {CSI},'' \emph{IEEE Trans.
  Signal Process.}, vol.~64, no.~12, pp. 3237--3249, June 2016.

\bibitem{Coordinateupdate}
F.~Sohrabi and T.~N. Davidson, ``Coordinate update algorithms for robust power
  loading for the {MU-MISO} downlink with outage constraints,'' \emph{IEEE
  Trans. Signal Process.}, vol.~64, no.~11, pp. 2761--2773, June 2016.

\bibitem{Arobustmaximin}
A.~Pascual-Iserte, D.~Palomar, A.~Perez-Neira, and M.~Lagunas, ``A robust
  maxmin approach for {MIMO} communications with imperfect channel channel
  state information based on convex optimization,'' \emph{{IEEE} Trans. Signal
  Process.}, vol.~45, pp. 346--360, Jan. 2006.

\bibitem{ATractableMethod}
N.~Vucic and H.~Boche, ``A tractable method for chance-constrained power
  control in downlink multiuser {MISO} systems with channel uncertainty,''
  \emph{IEEE Signal Proc. Letters}, vol.~16, no.~5, pp. 346--349, May 2009.

\bibitem{Reference4}
A.~Ben-Tal, L.~El~Ghaoui, and A.~Nemirovski, \emph{Robust optimization}.\hskip
  1em plus 0.5em minus 0.4em\relax Princeton University Press, 2009.

\bibitem{OntheDistributionof}
T.~Y. Al-Naffouri, M.~Moinuddin, N.~Ajeeb, B.~Hassibi, and A.~L. Moustakas,
  ``On the distribution of indefinite quadratic forms in {G}aussian random
  variables,'' \emph{IEEE Trans. Commun.}, vol.~64, no.~1, pp. 153--165, Jan.
  2016.

\bibitem{MIMObroadcast}
N.~Jindal, ``{MIMO} broadcast channels with finite-rate feedback,'' \emph{IEEE
  Trans.\ Inf.\ Theory}, vol.~52, no.~11, pp. 5045--5060, Nov. 2006.

\bibitem{cvx}
M.~Grant, S.~Boyd, and Y.~Ye, ``{CVX}: Matlab software for disciplined convex
  programming,'' 2008.

\bibitem{RobustSINR}
E.~Song, Q.~Shi, M.~Sanjabi, R.-Y. Sun, and Z.-Q. Luo,
  ``\BIBforeignlanguage{English}{Robust {SINR}-constrained {MISO} downlink
  beamforming: {W}hen is semidefinite programming relaxation tight?}''
  \emph{\BIBforeignlanguage{English}{EURASIP J. Wireless Commun. Networking}},
  vol.~1, pp. 1--11, 2012.

\bibitem{UnravelingtheRankOne}
W.~K. Ma, J.~Pan, A.~M.~C. So, and T.~H. Chang, ``Unraveling the rank-one
  solution mystery of robust {MISO} downlink transmit optimization: A
  verifiable sufficient condition via a new duality result,'' \emph{IEEE Trans.
  Signal Process.}, vol.~65, no.~7, pp. 1909--1924, April 2017.

\bibitem{Zeroforcingmethods}
Q.~H. Spencer, A.~L. Swindlehurst, and M.~Haardt, ``Zero-forcing methods for
  downlink spatial multiplexing in multiuser {MIMO} channels,'' \emph{{IEEE}
  Trans. Signal Process.}, vol.~52, no.~2, pp. 461--471, Feb. 2004.

\bibitem{Multipleantennachannelhardening}
B.~M. Hochwald, T.~L. Marzetta, and V.~Tarokh, ``Multiple-antenna channel
  hardening and its implications for rate feedback and scheduling,'' \emph{IEEE
  Trans. Inf. Theory}, vol.~50, no.~9, pp. 1893--1909, Sept. 2004.

\bibitem{Ontheoptimalityofmultiantenna}
T.~Yoo and A.~Goldsmith, ``On the optimality of multiantenna broadcast
  scheduling using zero-forcing beamforming,'' \emph{{IEEE} J. Sel. Topics
  Signal Process.}, vol.~24, no.~3, pp. 528--541, March 2006.

\end{thebibliography}
\end{document}